\documentclass[preprint,aps]{revtex4}
\usepackage{hyperref,slashed,color}
\usepackage{graphicx,subfig}
\usepackage{epstopdf}
\usepackage{multirow}

\begin{document}
\title{$Z'$ phenomenology: a model-independent analysis and fit in combination of chiral effective theory and anomaly cancellation }

\author{Ying Zhang\footnote{{\it Email address}: hepzhy@mail.xjtu.edu.cn},
Zheng Cai\footnote{{\it Email address}: caizheng.0@stu.xjtu.edu.cn}}

\address{School of Science, Xi'an Jiaotong University, Xi'an, 710049, P.R.China}
\date{Oct 26, 2011}                                           

\begin{abstract}
To investigate new gauge boson $Z'$ phenomenology model-independent, we combine chiral effective theory with anomaly cancellation conditions without any other model input. We focus on $Z'$ mixings with $\gamma-Z$ in both mass and kinetic parts and calculate contributions to oblique $S,T,U$.
The three sets of anomaly-free fermion $U(1)'$ charges parameterize the $Z'$ interactions with fermions.
The cancellation of the $[U(1)']^3$ anomaly and mixing gravitational-gauge anomaly determines the number of right-handed neutrinos.
We also find a novel relation between the charge assignments and Stueckelberg coupling in terms of the renormalized electromagnetic current. A global fit to the electroweak precise observables shows that typical values for the mixing parameters are of order $10^{-3}$. In spite of this strict limit, we obtain a negative $S$ parameter contribution.

\bigskip
PACS numbers: 12.60.Cn; 14.70.Pw; 12.15 Lk

\bigskip
Key words: new gauge boson $Z'$; anomaly free; chiral effective theory; fit
\end{abstract}
\maketitle

\section{introduction}\label{SECinctro}
Many puzzles in the standard model (SM) have prompted theorists to look for new physics by extending the SM. One introduces larger gauge groups
and more new particles to try to answer problems existing in the SM. A familiar and general characteristic of new physics is extending the Abelian gauge group associated with extra neutral vector bosons, usually labeled by $Z'$. $Z'$ is often the lightest new particles beyond SM and easier to find in new colliders. Another reason is that $Z'$ may play many important roles in theory, such as mediating the hidden sector, breaking SUSY, and solving the $\mu$ problem in minimal supersymmetric standard model (MSSM) \cite{LangackerArxiv2008,LangackerPRL2008}.

	There are two issues arising from the new vector boson that pique our interests. One is $Z'$ mixings with electroweak neutral bosons $Z$ and $\gamma$, which is a fashionable means to affect low-energy scale physics. This translates into  very high sensitivity for  electroweak precise observables (EWPO) that can be performed at  $Z$ resonance. Many authors have investigated the issue and provided bounds on $Z'$ \cite{LangackerPRD1992,KunduPLB1996,ErlerPRL2000,LeikePR1999}. Usually, a lighter $Z'$ is possible for larger mixing angles, although \ smaller mixing angles  exist only for heavier $Z'$ \cite{HewettPR1989}.	
	More detailed results depend on mixing forms set by the models. With the Exception of the minimal $Z-Z'$ mass mixing, $Z-Z'$ kinetic mixing is also discussed \cite{AbelJHEP2008, FeldmanPRD2007,BabuPRD1998,KumarPRD2006} motivated by enlargement of the parameter space. Because  gauge symmetry allows the  existence of kinetic mixings, we should consider all possible kinetic mixings despite their  complexity.   Other motivations come from  special applications in super-gravity and string theory models \cite{DienesNPB1997}. The number of  mixing parameters needed to describe complete mixings is the first question we will resolve in this paper.
	
	Another interesting issue is $Z'$ interactions with leptons and quarks. As is well known, for a given $U(1)'$ gauge coupling $g''$, $Z'$ interactions are decided by charges assignment to fermions. In different models, $U(1)'$ charges are assigned according to different consideration. Phenomenological results are as a consequence highly model-dependent.  In theory, new gauge group charge assignments must cancel the anomaly in the triangle diagrams to maintain gauge invariance. We study all anomaly cancellation conditions to find the anomaly-free solutions wheher considering right-handed neutrinos or not. 

We investigate in a model-independent manner the general mixings and interactions of extra neutral gauge boson in the spirit of Weinberg's effective lagrangian and anomaly cancellation conditions to find a most probable parameters space. 
In Sec.\ref{SECmixing}, we reivew the most general mixings, including mass mixings and kinetic mixings. A three-body rotation matrix with Weinberg angle and $Z'$ correction terms is introduced to diagonalize the mixing matrices. These rotation matrix elements stands for new physics effect and are decided by underlying chiral effective Lagrangian corresponding to extra electroweak symmetry $SU(2)_L\otimes  U(1)_Y\otimes U(1)'$. We calculate oblique radiative corrections from the similar to Holdom \cite{HoldomPLB1991}. In Sec.\ref{SECcharge},  we discuss $Z'$ interactions with leptons and quarks. We find that vector-type electromagnetic coupling yields a constraint on rotation matrix elements and disregards Stueckelberg coupling unless the $Z'$ coupling fermion as in the $B-L$ model and right-handed neutrinos $\nu_R$ are involved. Couplings correction from gauge boson mixings is derived which results depend on mixing parameters and fermion $U(1)'$ charges. Instead of model inputting charges,  we assign $U(1)'$ charges to the fermions with the requirement that these cancel the gauge anomaly in the triangle diagrams. We find that the number of $\nu_R$ is decided by $[U(1)']^3$ and mixing gravitational-gauge anomaly cancellation conditions. With anomaly-free charge assignments, $Z'$ phenomenology is studied. Branching ratio $R'_{l,q}$ and asymmetry $A'_{l,q}$ of $Z'$ only depend on the single charge ratio $y'_q/y'_u$ for the light fermion case. Also, a $Z'$ effect at the low energy scale can be fit to EWPO. In Sec.\ref{SECfit},  we explain the the fitting method and aspects of the set-up, results of which are listed in Sec.\ref{SECconclusion}. Using these results, oblique radiative corrections $S,T,U$ are shown to be bound. In particular, the $S$ parameter can take negative values within $0.95$ CL.
\section{Neutral bosons mixings}\label{SECmixing}
\subsection{General mixing inspired by effective theory}	
As  mentioned above, mixing involves processes by which extra neutral gauge bosons $Z'$ effect  low-energy scale physics. The simplest mixing is  minimal $Z$-$Z'$ mass mixing. However, single parameter mass mixing is not enough to describe all possible $Z'$ physics. Kinetic mixing should be introduced although that increases the region of parameter space. Moreover, kinetic mixing often exists in a broad class of supergravity and string models \cite{AbelJHEP2008,FeldmanPRD2007}. In the section, we will review a general three-body mixing corresponding to $Z$-$A$-$Z'$ in both mass and kinetic parts in terms of chiral effective theory constructed in our early works \cite{OurJHEP2008,OurJHEP2009}.
All possible $Z'$ mixing terms in both mass and kinetic parts can be set into 
\begin{eqnarray}
\mathcal{L}&=&
-\frac{1}{4}f^2\mathrm{tr}[\hat{V}_\mu\hat{V}^\mu]
+\frac{1}{4}\beta f^2\mathrm{tr}[\hat{V}_\mu]\mathrm{tr}[T\hat{V}^\mu]
+\frac{1}{2}\alpha_aB_{\mu\nu}\mathrm{tr}[TW^{\mu\nu}]
\nonumber\\
&&+\frac{1}{4}\alpha_b\mathrm{tr}[TW_{\mu\nu}]\mathrm{tr}[TW^{\mu\nu}]
+\alpha_cX_{\mu\nu}\mathrm{tr}[TW^{\mu\nu}]
+\alpha_dB_{\mu\nu}X^{\mu\nu}
\label{Lagrangian}
\end{eqnarray}
Here, $W_\mu$, $B_\mu$ and $X_\mu$ are $SU(2)_L$, $U(1)_Y$ and $U(1)'$ gauge fields, respectively.
However, $T=\hat{U}\tau_3\hat{U}^\dag$ and $\hat{V}=(\hat{D}_\mu\hat{U})\hat{U}^\dag$ are $SU(2)_L$ covariant operators with Goldstone bosons non-linear realization $\hat{U}$ and covariant derivation $\hat{D}_\mu\hat{U}$ defined as
$$\hat{D}_\mu\hat{U}=\partial_\mu\hat{U}+igW_\mu\hat{U}-ig'\hat{U}\frac{\tau_3}{2}B_\mu-i\hat{U}(\tilde{g}'B_\mu+g''X_\mu)I.$$
Note that we have introduced  Stueckelberg coupling $\tilde{g}'$ to generate partly $Z'$ mass by the Stueckelberg mechanism \cite{KorsJHEP2005}. 
The first term in (\ref{Lagrangian}) is non-linear $\sigma$ model.
Although, there are other mass terms, such as $(\mathrm{tr}[T\hat{V}_\mu])^2$ and $(\mathrm{tr}[\hat{V}_\mu])^2$, to invoke mass mixing, they can be absorbed into redefinitions of the electroweak gauge couplings $g,g'$ \cite{OurarXiv2010}. 
Taken unitary gauge $\hat{U}=1$ corresponding to electroweak symmetry breaking, mass square matrix and kinetic matrix become
\begin{eqnarray}
\mathcal{M}_0^2&=&f^2\left(\begin{array}{ccc} \frac{g^2}{4}&-\frac{gg'}{4}+\frac{g\tilde{g}'}{2}\beta
&\frac{gg''}{2}\beta
\\ -\frac{gg'}{4}+\frac{g\tilde{g}'}{2}\beta&\frac{g'^2}{4}+\tilde{g}'^2-g\tilde{g}'\beta
&-\frac{g'g''}{2}\beta+g''\tilde{g}'\\
	\frac{gg''}{2}\beta&-\frac{g'g''}{2}\beta+g''\tilde{g}'
&g''^2
	\end{array}\right)\;,\\
\mathcal{K}_0&=&-\frac{1}{4}\left(\begin{array}{ccc}
		1-\alpha_b&-\alpha_a
&-2\alpha_c	\\
		-\alpha_a&1
&-2\alpha_d
		\\
		-2\alpha_c&-2\alpha_d&1		
	\end{array}\right)\;.
\end{eqnarray}
Here, $\alpha_{a,b,c,d}$ control the kinetic mixing, $\beta$ is a single mass mixing parameter, and the Stueckelberg coupling $\tilde{g}'$ also yields mass mixing.  
\subsection{Diagonalization} 
To diagonalize $\mathcal{M}_0^2$ and $\mathcal{K}_0$ simultaneously, we need nine independent parameters corresponding to four kinetic mixing $\alpha_{abcd}$, one mass mixing $\beta$, one Stueckelberg coupling $\tilde{g}'$, two  gauge coupling ratios  ${g_Z}/{g''}$ and ${g}/{g'}$, and one normalized photon factor. Generally, the rotation matrix between gauge eigenstates and mass eigenstates can be written 
\begin{eqnarray}
	\left(\begin{array}{c}W^3_\mu\\ B_\mu\\ X_\mu\end{array}\right)
		=U\left(\begin{array}{c}Z_\mu\\ A_\mu\\ Z'_\mu\end{array}\right)
\label{Urotation}
\end{eqnarray}
with
\begin{eqnarray}
	U&=&\left(\begin{array}{ccc}
		U_{11}&  U_{12} &   U_{13}
		\\
		 U_{21} &  U_{22} & U_{23}
		\\
		U_{31} & U_{32} & U_{33}
		\end{array}\right)
		=\left(\begin{array}{ccc}
		c_W+\Delta_{11}& s_W+\Delta_{12} &  \Delta_{13}
		\\
		-s_W+\Delta_{21} & c_W+\Delta_{22} & \Delta_{23}
		\\
		\Delta_{31} & \Delta_{32} &1+ \Delta_{33}
		\end{array}\right)
\label{Umatrix}
\end{eqnarray}
In the above, we write $U$ as a standard electroweak rotation adding to nine mixing contributions $\Delta_{ij}$. Here $U$ includes ten parameters. To match only nine independent parameters, we must find a constraint relation between the various $\Delta_{ij}$ which will arise from renormalized electromagnetic currents that relate to  $\Delta_{12}$ and $\Delta_{22}$. We perform this in the next section. 

The rotation matrix $U$ can be determined by the underlying chiral effective theory to satisfy
\begin{eqnarray}
	U^T\mathcal{M}_0^2U={\bf diag}(M_Z^2,0,M_{Z'}^2),~~~~
	U^T\mathcal{K}_0U=-\frac{1}{4}I.
\label{Ueq}
\end{eqnarray}
The detail formulae are listed in Appendix \ref{APProtationmatrix}. If all nine independent parameters vanish, i.e. $\Delta_{ij}=0$, $U$ reduces to the standard electroweak rotation. Due to the success of fitting the SM to experiment data, we believe $\Delta_{ij}$ should be small enough that there should be slight shifts in the electroweak observables. 

After rotating the $U$ matrix, neutral gauge fields would then be diagonalized to mass eigenstates.   The $Z$ and $Z'$  masses are read from (\ref{Lagrangian}) and (\ref{Umatrix}) as
\begin{eqnarray}
	M_Z^2
	&=&f^2\left\{\frac{1}{2}
	\left(\frac{e}{c_Ws_W}+(\frac{e}{s_W}\Delta_{11}-\frac{e}{c_W}\Delta_{21})\right)^2+g''^2\Delta_{31}^2\right\}
	\label{ZMass}\\
	M_{Z'}^2&=&f^2\left\{g''^2(1+\Delta_{33})^2+\frac{1}{4}(g\Delta_{13}-g'\Delta_{23})^2\right\}
	\label{ZprimeMass}
\end{eqnarray}
with the SM mass of $Z$ $M_{Z_0}=\frac{ef}{2c_Ws_W}$.

\subsection{Oblique radiative corrections}
Non-standard mixings of electroweak neutral gauge bosons will directly shift oblique radiative corrections $S,T,U$. Using Holdom's procedure \cite{HoldomPLB1991}, we can calculate the $Z'$ corrections to $S,T,U$ as follows
\begin{eqnarray}
\Delta S&=&\frac{4s_Wc_W}{\alpha}\left\{(s_W\Delta_{11}-2s_Wc_W(s_W\Delta_{12}+c_W\Delta_{22})-c_W\Delta_{21})\right\}
\label{S}\\
\Delta T&\simeq&-\frac{4s_W^2c_W^2}{\alpha}\frac{g''^2}{e^2}\Delta_{31}^2
\label{T}\\
\Delta U&=&-\frac{8s_W^2}{\alpha}(c_W\Delta_{11}+s_W^3\Delta_{12}+s_W^2c_W\Delta_{22})
\label{U}
\end{eqnarray}
These formulae  agree with those of Appelquist which express the corrections in terms of the coefficients of electroweak chiral Lagrangian in \cite{AppelquistPRD2003}.

\section{Fermions interaction}\label{SECcharge}
In this section, we discuss the $Z'$ interaction with fermions. Given fixed $U(1)'$ gauge coupling $g''$, the $Z'$ interactions with leptons and quarks are dominated by the  $U(1)'$ fermion charges. Initially, neutral current interactions are investigated and the corrections to vector and axial-vector couplings is derived from the $Z'$ mixing. We then study the charge assignments according to the anomaly cancellation conditions and give all possible  anomaly-free solutions. Moreover, we calculate  $Z'$ decay  under all kinds of  anomaly-free charge assignment.
\subsection{Extra neutral current}
 Neutral current interactions  including one extra $Z'$ boson in Lagrangian are
\begin{eqnarray}
-\mathcal{L}_{NC}&=&gW^3_\mu J^{3,\mu}+g'B_\mu J^\mu_Y+{g''}X_\mu J^\mu_X.
\nonumber
\end{eqnarray}
Here,
\begin{eqnarray}
J^{\mu}_3&=&\sum_i\bar{f}_i\gamma^\mu t_{3iL}P_Lf_i
\nonumber\\
J^\mu_Y&=&\sum_i\bar{f}_i\gamma^\mu[y_{iL}P_L+y_{iR}P_R]f_i
\nonumber\\
J^\mu_X&=&\sum_i\bar{f}_i\gamma^\mu[y'_{iL}P_L+y'_{iR}P_R]f_i
\nonumber
\end{eqnarray}
are neutral currents corresponding to weak isospin third component $W^3_\mu$, hypercharge $B_\mu$ and extra $U(1)'$ boson $X_\mu$, respectively. $y'_{iL,R}$ is the left/right-handed  fermionic $U(1)'$ charges with flavor index $i$.
After spontaneous breaking $SU(2)\otimes U(1)_Y\otimes U(1)'$ to the electromagnetic
subgroup $U(1)_{em}$, $Z$ and $Z'$ obtain masses while maintaining the photon massless.  In
the mass eigenstates basis, neutral currents become
\begin{eqnarray}
-\mathcal{L}_{NC}&=&e^*J_{em}^\mu A_\mu+g_ZJ_Z^\mu Z_\mu+{g''}J_{Z'}^\mu Z'_\mu.
\nonumber
\end{eqnarray}
Here,
\begin{eqnarray}
e^*J_{em}^\mu
&=&e^*\sum_i\bar{f}_i\gamma^\mu q_i f_i
\nonumber\\
J_Z^\mu
&=&\sum_i\bar{f}_i\gamma^\mu(\epsilon_{iL}P_L+\epsilon_{iR}P_R)f_i
=\frac{1}{2}\sum_i\bar{f}_i\gamma^\mu(g_{iV}-g_{iA}\gamma_5)f_i
\nonumber\\
J_{Z'}^\mu&=&\sum_i\bar{f}_i\gamma^\mu(\epsilon'_{iL}P_L+\epsilon'_{iR}P_R)f_i
=\frac{1}{2}\sum_i\bar{f}_i\gamma^\mu(g'_{iV}-g'_{iA}\gamma_5)f_i
\nonumber
\end{eqnarray}
are currents corresponding to electromagnetic, $Z$ and $Z'$, respectively; 
the vector and axial-vector couplings are $g_{iV,A}=\epsilon_{iL}\pm\epsilon_{iR}$, $g'_{iV,A}=\epsilon'_{iL}\pm\epsilon'_{iR}$;
and $e^*$ is the renormalized electric change.
With the help of (\ref{Urotation}), we can read out 
\begin{eqnarray}
g_ZJ^\mu_Z&=&gU_{11}J^{3,\mu}+g'U_{21}J^{\mu}_Y+{g''}U_{31}J^\mu_X
\label{Jz}\\
e^*J_{em}^\mu&=&gU_{12}J^{3,\mu}+g'U_{22}J^{\mu}_Y+{g''}U_{32}J^\mu_X
\label{Jem}\\
{g''}J_{Z'}^\mu&=&gU_{13}J^{3,\mu}+g'U_{23}J^{\mu}_Y+{g''}U_{33}J^\mu_X
\label{Jzprime}
\end{eqnarray}
The renormalized electric charge is
	\begin{eqnarray}
	e^*q_i
	=eq_i
		+\frac{e}{c_W}\Delta_{22}[y_{iL}P_L+y_{iR}P_R]+\frac{e}{s_W}\Delta_{12} t_{3iL}P_L+{g''}\Delta_{32}[y'_{iL}P_L+y'_{iR}P_R].
	\label{electriccharge}
	\end{eqnarray}
Note that experimentally the electromagnetic coupling is vector-type that requires an equal coupling of the left-handed  to the right-handed  eigenstates.
We obtain  two constraint conditions on mixings and charges:
	\begin{itemize}
		\item[1.]  $s_W\Delta_{22}=c_W\Delta_{12}$. This constraint can be expressed in terms of a  rotation matrix in Appendix \ref{APProtationmatrix}.  In the particle physics context, it arises from the requirement for a massless photon. In the gauge eigenstate basis, the factor $c_WW^3_\mu-s_WB_\mu$ generates a weak boson $Z$ mass, that could include some component of a massive $Z'$ by $Z-Z'$ mixing. 		
		\begin{eqnarray*}
			c_WW^3_\mu-s_WB_\mu&=&(c_WU_{11}-s_WU_{21})Z_\mu
			\\
			&&+(c_WU_{12}-s_WU_{22})A_\mu+(c_WU_{13}-s_WU_{23})Z'
		\end{eqnarray*}
		However, it is forbidden to contain any component of photon so that the photon remains massless. 
Letting the second term vanish on the r.h.s. of the  above equation, we obtain
		\begin{eqnarray*}
			0=c_WU_{12}-s_WU_{22}=c_W\Delta_{12}-s_W\Delta_{22}.
		\end{eqnarray*}
		providing one constraint condition.
		\item[2.] Another constraint comes from the last term in (\ref{electriccharge}). Here, we have two auxiliary choices: either a vanishing $\Delta_{32}$ or $y'_{iL}=y'_{iR}$ for each flavor. We know that $\Delta_{32}$ in the rotation matrix $U$ plays a role in diagonalizing the Stueckelberg mixing \cite{OurJHEP2008}. Thus, the former choice means that the $\tilde{g}'$ vanishes. The latter strictly limits the $Z'$ interaction with fermions. In particular, in next subsection, we will see that the latter case yields $B-L$ type anomaly-free charge assignments.
	\end{itemize}
For this reason, it is surprising that  the total left- and right-handed couplings are not required to be equal in (\ref{electriccharge}), yielding a single constraint. The reason is that, for a very heavy $Z'$, the $Z'$ mixing tends to vanish and the total constraint reduces to  the first constraint $s_W\Delta_{22}=c_W\Delta_{12}$. From another point of view, the first constraint will arise from a $2\times2$ mixing space without $U(1)'$. When adding $U(1)'$ into the electroweak group, the second constraint appears. In this way, we can say that the two constraints do not include any enhancement.

$Z'$ mixing makes the vector  $g_V$ and axial-vector $g_A$ couplings diverge from  SM values.
  From (\ref{Jz}) and (\ref{Jzprime}), we can obtain coupling corrections 
\begin{eqnarray}
	\delta g_{iV}
		&=&c_W\Delta_{11} t_{3iL}+s_W\Delta_{21}(y_{iL}+y_{iR})+\frac{g''s_Wc_W}{e}\Delta_{31}(y'_{iL}+y'_{iR})
	\label{deltagv}\\
	\delta g_{iA}
		&=&c_W\Delta_{11} t_{3iL}+s_W\Delta_{21}(y_{iL}-y_{iR})+\frac{g''s_Wc_W}{e}\Delta_{31}(y'_{iL}-y'_{iR})
	\label{deltaga}
\end{eqnarray}
which provide $Z'$ low energy corrections. Phenomenologically, these shifts will correct EWPO and can be bounded by electroweak precise test (in Sec. \ref{SECfit}, we will return to this issue).  

In contrast, neutral bosons mixings also correct $Z'$ couplings to fermions as follows
\begin{eqnarray}
	\delta g'_{iV}&=&\frac{g}{g''}\Delta_{13}t_{3iL}+\frac{g'}{g''}\Delta_{23}(y_{iL}+y_{iR})+\Delta_{33} (y'_{iL}+y'_{iR})
	\label{deltagvprime}\\
	\delta g'_{iA}&=&\frac{g}{g''}\Delta_{13}t_{3iL}+\frac{g'}{g''}\Delta_{23}(y_{iL}-y_{iR})+\Delta_{33} (y'_{iL}-y'_{iR})
	\label{deltagaprime}
\end{eqnarray}
For a heavy $Z'$ , mixing corrections usually are regards as  negligible, which means $Z'$ couplings are mainly determined by the fermion charge $y'_{iL,R}$.
\subsection{Charge assignment}
Up to now, $U(1)'$ fermion charges $y'_{iL,R}$ are random parameters that take different values in many $Z'$ models. 
Except for the SM fermions, we always regard the right-handed neutrino as an exotic fermion in new physics models. 
Usually, fermions are assigned universal family charges to avoid issues from  flavor changing neutral currents.
However, in the  model-independent case, 
we are interested in the number of independent parameters is needed to describe universal family charge assignments.
For example, in an $SU(2)_L$ symmetry, we can assign the same $U(1)'$ charge to the left-handed fermion two components, i.e.
$y'_{u_L}=y'_{d_L}\equiv y'_q$ for quarks and
$y'_{\nu_L}=y'_{e_L}\equiv y'_l$ for leptons. Thus, the  universal family charge assignment can be described by six charge parameters:
$y'_l$ and $y'_q$ for both left-handed leptons and quarks, $y'_u$, $y'_d$, $y'_e$ and $y'_{\nu_R}$ for the right-handed up quark,
right-handed down quark, right-handed electron and right-handed neutrino, respectively.

Although the $Z'$ charge ca not be determined at this stage by current experiments,
the $U(1)'$ charges of quarks and leptons must cancel the triangular anomaly to
preserve gauge symmetry in the theory \cite{DavidsonPRD1979,CarenaPRD2004,AppelquistPRD2003}. The anomaly cancellation conditions 
can reduce the number of free charges to improve prediction of theory. Additionally, we will prove below that the number of right-handed neutrinos is three or zero to cancel separately the $[U(1)']^3$ anomaly and the mixed gravitational-gauge anomaly.

The anomaly cancellation conditions for SM fermions (no right-handed neutrino) include $[SU(3)_C]^2U(1)'$, $[SU(2)_L]^2U(1)'$, $U(1)_Y[U(1)']^2$, and $[U(1)_Y]^2U(1)'$ anomalies, which require that
\begin{eqnarray}
\left\{\begin{array}{l}
2y'_q -y'_d-y'_u=0
\\
y'_l+3y'_q=0
\\
-{y'_l}^2+{y'_q}^2+{y'_e}^2-2{y'_u}^2+{y'_d}^2=0
\\
3y'_l+y'_q-6y'_e-8y'_u-2y'_d=0
\end{array}\right.
\end{eqnarray}
Solving the above four equations, we find that the charge assignments for the  SM fermions is parameterized by two free charges
\begin{eqnarray}
    y'_l=-3y'_q,~~~
    y'_d=2y'_q-y'_u,~~~
    y'_e=-2y'_q-y'_u.
\label{ChargeAssignmentI}
\end{eqnarray}

Next, let us now consider the  right-handed neutrinos.
There are two anomaly cancellation conditions, viz. the $[U(1)']^3$ anomaly and the  mixing gravitational-gauge anomaly,  associated to $y'_{\nu_R}$
\begin{eqnarray}
\left\{\begin{array}{c}
2{y'^3_l}+6{y'^3_q}-{y'^3_e}-3{y'^3_u}-3{y'^3_d}-Ny'^3_{\nu_R}=0
\\
3y'_q+y'_l-3y'_u-3y'_d-y'_e-Ny'_{\nu_R}=0
\\
\end{array}\right.
\end{eqnarray}
Here, $N$ is the number of right-handed neutrinos for each generation. Substituting (\ref{ChargeAssignmentI}), a solution for $y'_{\nu_R}$ 
exists if and only if $N=1$.
\begin{eqnarray}
y'_{\nu_R}=y'_u-4y'_q
\label{ChargeAssignmentII}
\end{eqnarray}
This implies that there is only one right-handed neutrino for each generation. 
If we relax  the constraint requiring  two parameter dependence in (\ref{ChargeAssignmentI}) to allow $y'_u=4y'_q$, a right-handed neutrino may not exist (or does not couple to $Z'$). In that case,  charge assignments of the SM fermions are described by only one free charge 
\begin{eqnarray}
    y'_u=4y'_q,~~~
    y'_l=-3y'_q,~~~
    y'_d=-2y'_q,~~~
    y'_e=-6y'_q.
\label{SMChargeAssignment}
\end{eqnarray}

Notice that if we use the Stueckelberg coupling $g''$ to choose $y'_{i,L}=y'_{i,R}$ for leptons and quarks, three right-handed neutrinos must exist to avoid a trivial solution where all couplings vanish. We can say the Stueckelberg coupling 'loves' right-handed neutrinos. At this point, the anomaly cancelling charge assignments become $B-L$ type, i.e. 
$$y'_u=y'_d=y'_q=1/3,~~~y'_e=y'_l=-1$$ 
with a random constant of proportionality. 

In brief, according to the renormalized electric charge (\ref{electriccharge}) and existence of $\nu_R$, there are three kinds of anomaly-free charges:
	\begin{itemize}
		\item case 1: Stueckelberg coupling $\tilde{g}'=0$  and the existence of three $\nu_R$. The mixing matrix element $\Delta_{32}$ is equal to zero; the charge assignments are controlled by two free charges: $y'_q$ and $y'_u$. 
		\item case 2: Stueckelberg coupling $g''=0$ and no $\nu_R$. Here,  $\Delta_{32}$ vanishes and the charge assignments are controlled by single free charge. 
		\item case 3: Stueckelberg coupling $g''\neq0$.  Three $\nu_R$ must exist to preserve the anomaly cancelling solution; the charge assignments are $B-L$ type.
	\end{itemize}

\subsection{$Z'$ decay}
We use the above charge assignments to predict decay the process of $Z'$ to a fermion pair, a decay that up- coming collider experiments can measure. Its couplings to fermions determine the leading $Z'$ decay.
Negslecting mixing corrections, the decay width $\Gamma_{Z'}(f\bar{f})$ for a massless fermion pair $f\bar{f}$ is given by
\begin{eqnarray*}
\Gamma_{Z'}(f\bar{f})
=\frac{N_fG_FM_{Z'}^3}{6\sqrt{2}\pi}({g'}_{iV}^2+{g'}_{iA}^2)
\end{eqnarray*}
For leptons $N_f=1$, whereas for quarks $N_f=3$. The vector and axial-vector couplings to $Z'$ are $g'_{iV,A}=y'_{iL}\pm y'_{iR}$.
As a matter of convenience, we express the decay widths in terms of $Z'$ charge $y'_{iL,iR}$
\begin{eqnarray*}
\Gamma_{Z'}(f\bar{f})
=\frac{N_fG_FM_{Z'}^3}{3\sqrt{2}\pi}\Big\{{y'}^2_{iL}+{y'}^2_{iR}\Big\}
\end{eqnarray*}
With the help of the anomaly cancelling solution, the $Z'$ decay width can be expressed using two independent parameters $y'_q$ and $y'_u$ (in case 1).
Now, let us discuss the decays under the anomaly  cancelling solutions (\ref{ChargeAssignmentI}) and (\ref{ChargeAssignmentII}).
The $Z'$ decay widths to different flavors are
\begin{eqnarray}
\begin{array}{rcl}
\Gamma_{Z'}(u\bar{u})
    &=&\frac{G_FM_{Z'}^3}{\sqrt{2}\pi}\Big\{{y'}^2_{q}+{y'}^2_{u}\Big\}
\\
\Gamma_{Z'}(d\bar{d})
    &=&\frac{G_FM_{Z'}^3}{\sqrt{2}\pi}\Big\{5{y'}^2_{q}-4y'_qy'_u+y'^2_u\Big\}
\\
\Gamma_{Z'}(e\bar{e})
    &=&\frac{G_FM_{Z'}^3}{3\sqrt{2}\pi}\Big\{13{y'}^2_{q}+4y'_qy'_u+y'^2_u\Big\}
\\
\Gamma_{Z'}(\nu\bar{\nu})
    &=&\frac{G_FM_{Z'}^3}{3\sqrt{2}\pi}\Big\{25{y'}^2_{q}-8y'_qy'_u+y'^2_u\Big\}
\end{array}\label{DecayWidthI}
\end{eqnarray}
The total $Z'$ decay width is determined by summing over all flavors $\Gamma_{Z'}=\sum_{f}\Gamma_{Z'}(f\bar{f})$
and the hadronic decay width by the summing over all quarks $\Gamma_{Z'}(had.)=\sum_{f=quarks}\Gamma_{Z'}(q\bar{q}).$
In particular, the hadron-to-lepton ratios $R'_e$, and $R'_{\nu}$ as well as the hadron branching ratio $R'_{b}$ and $R'_t$ are determined by only one free parameter, viz. the charge ratio  $r\equiv y'_q/y'_u$ (see Fig.1 for details)
\begin{figure}
\includegraphics[scale=0.5]{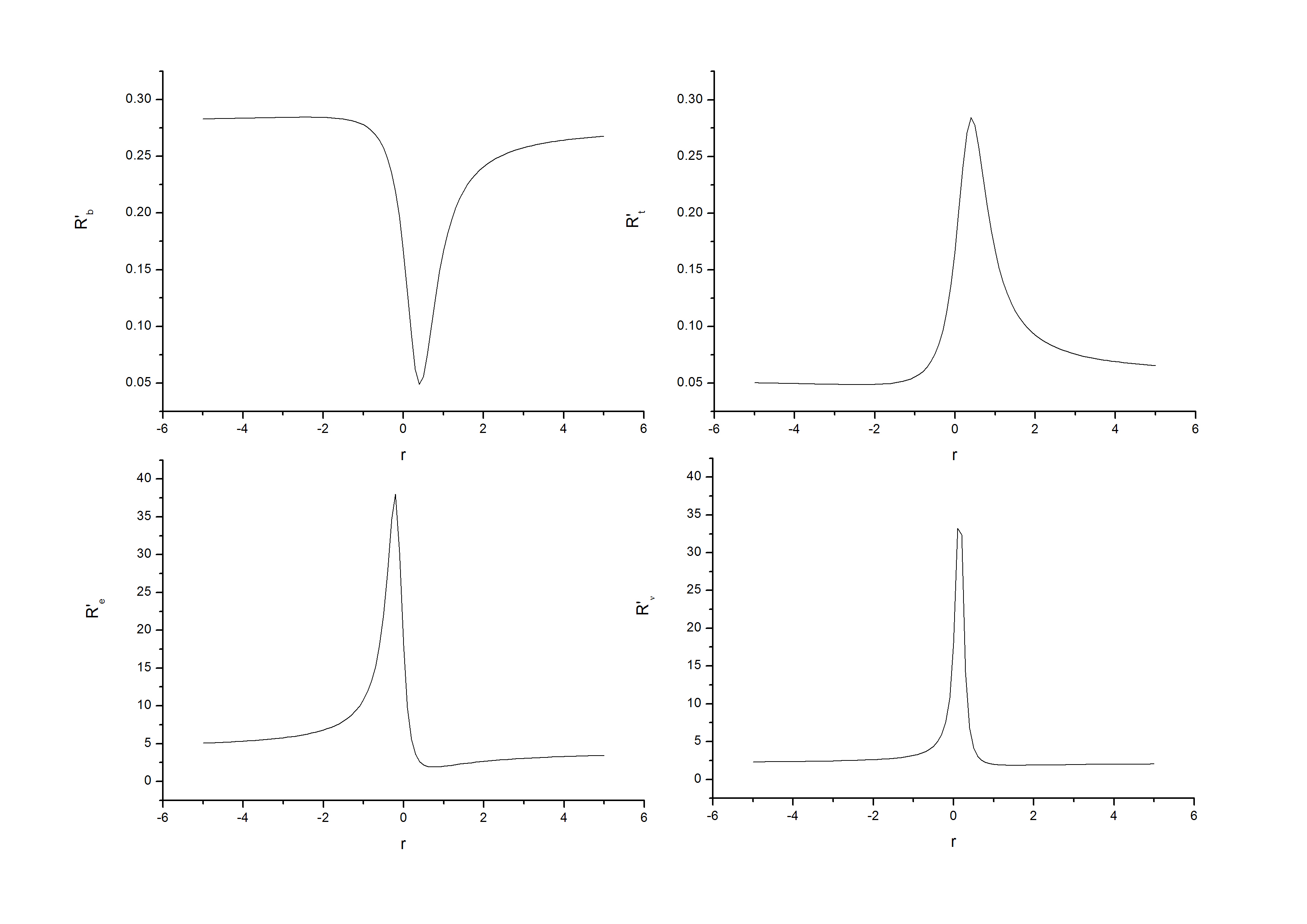}
\caption{$Z'$ branching ratios for massless fermions vs. $U(1)'$ charge ratio $r$}
\label{FIG:Rmassless}
\end{figure}

Furthermore, in considering the fermion mass corrections, the $Z'$ decay width to a fermion pair \cite{RobinettPRD1982} is 
\begin{eqnarray*}
\Gamma_{Z'}(f\bar{f})
=\frac{\mu N_fG_FM_{Z'}^3}{6\sqrt{2}\pi}\Big\{({g'}_{iV}^2+{g'}_{iA}^2)(1+\frac{2m_f^2}{M_{Z'}^2})-6g_{iA}^2\frac{m_f^2}{M_{Z'}^2}\Big\}
\end{eqnarray*}
with fermion mass $m_f$ and  phase space factor arising from the massive final fermions $\mu=\sqrt{1-4m_f^2/M_{Z'}^2}$. 
 Although, for a heavier top quark, the effect of fermion mass is more prominent than for  other quarks, yielding only a slight shift in Fig. 2.
\begin{figure}
\includegraphics[scale=0.25]{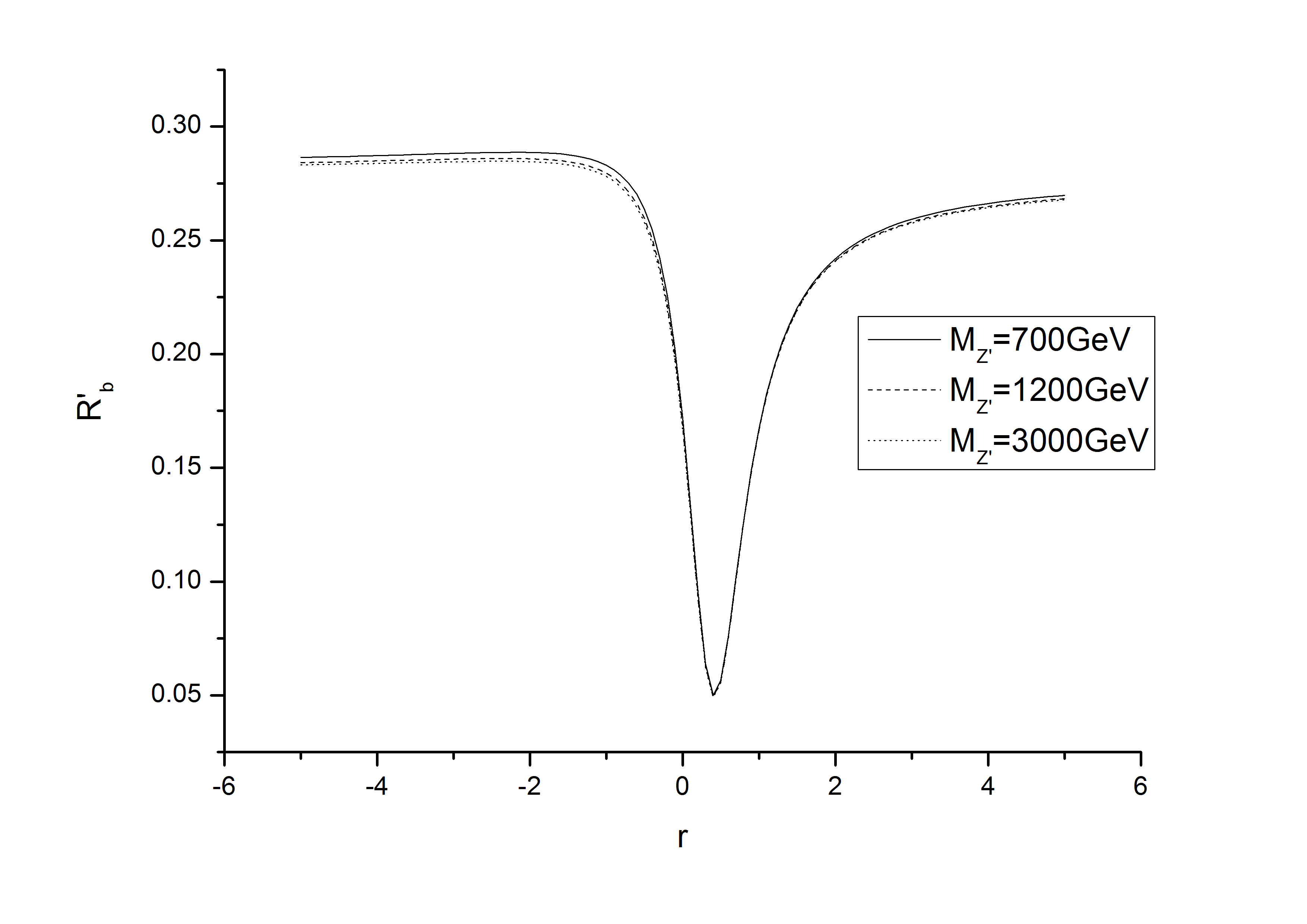}
\includegraphics[scale=0.25]{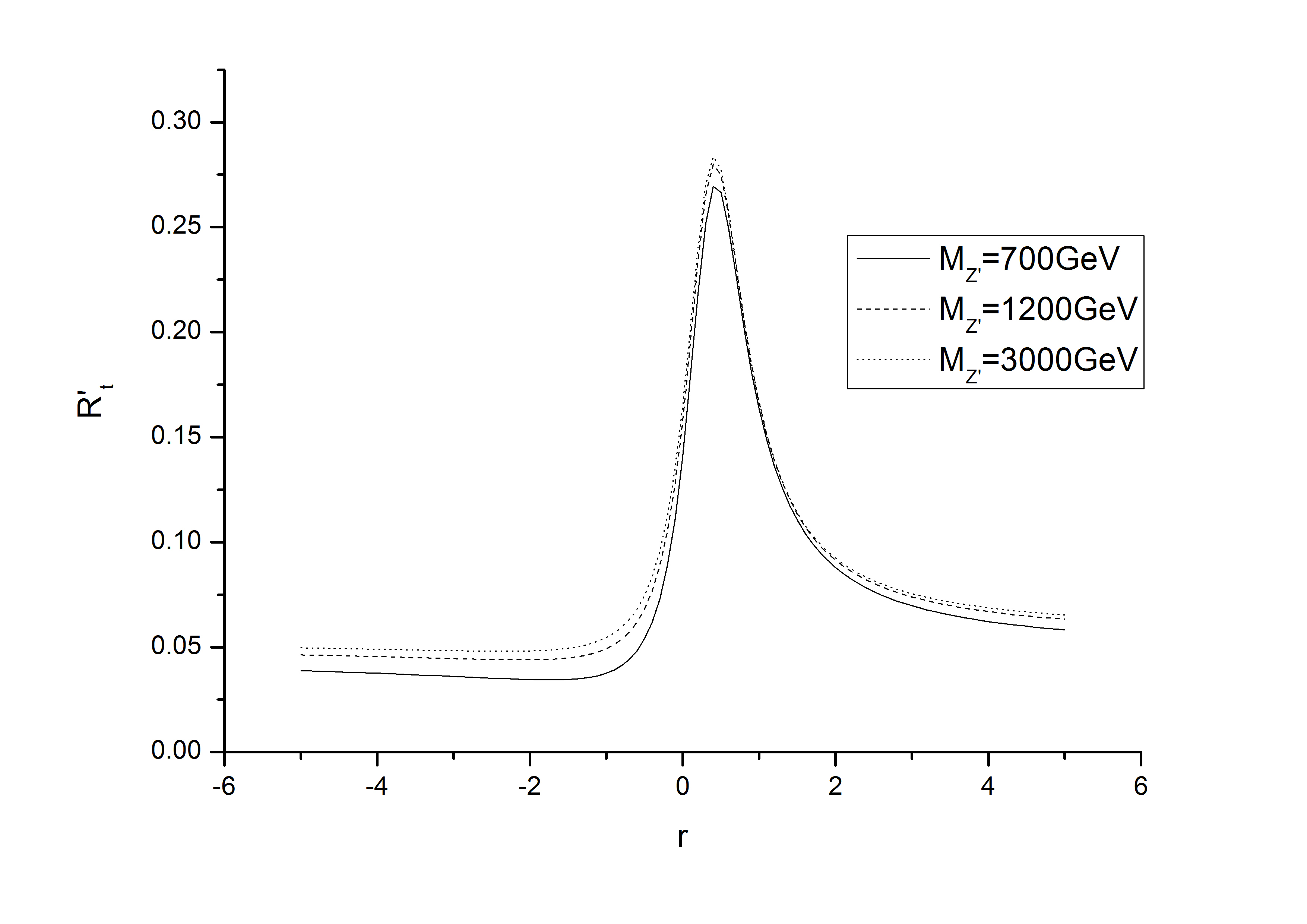}
\caption{$Z'$ branching ratios for heavy $b$ quark and $t$ quark vs. charge ratio $r$ at $M_{Z'}=700GeV, 1200GeV, 3000GeV$}
\label{Rmass}
\end{figure}

Similarly,  we can discuss the left-right asymmetry $A'_{LR}(f)$ and the forward-backward asymmetry $A'_{FB}(f)$ of $Z'$
\begin{eqnarray*}
A'_{LR}(f)&=&\frac{2g'_{V}g'_A}{{g'}_V^2+{g'}_A^2}
\\
A'_{FB}(f)&=&\frac{3}{4}A'_{LR}(e)A'_{LR}(f).
\end{eqnarray*}
These are determined from only charge ratio $r$ which we graph in Figs.3 and 4.
\begin{figure}
\includegraphics[scale=0.50]{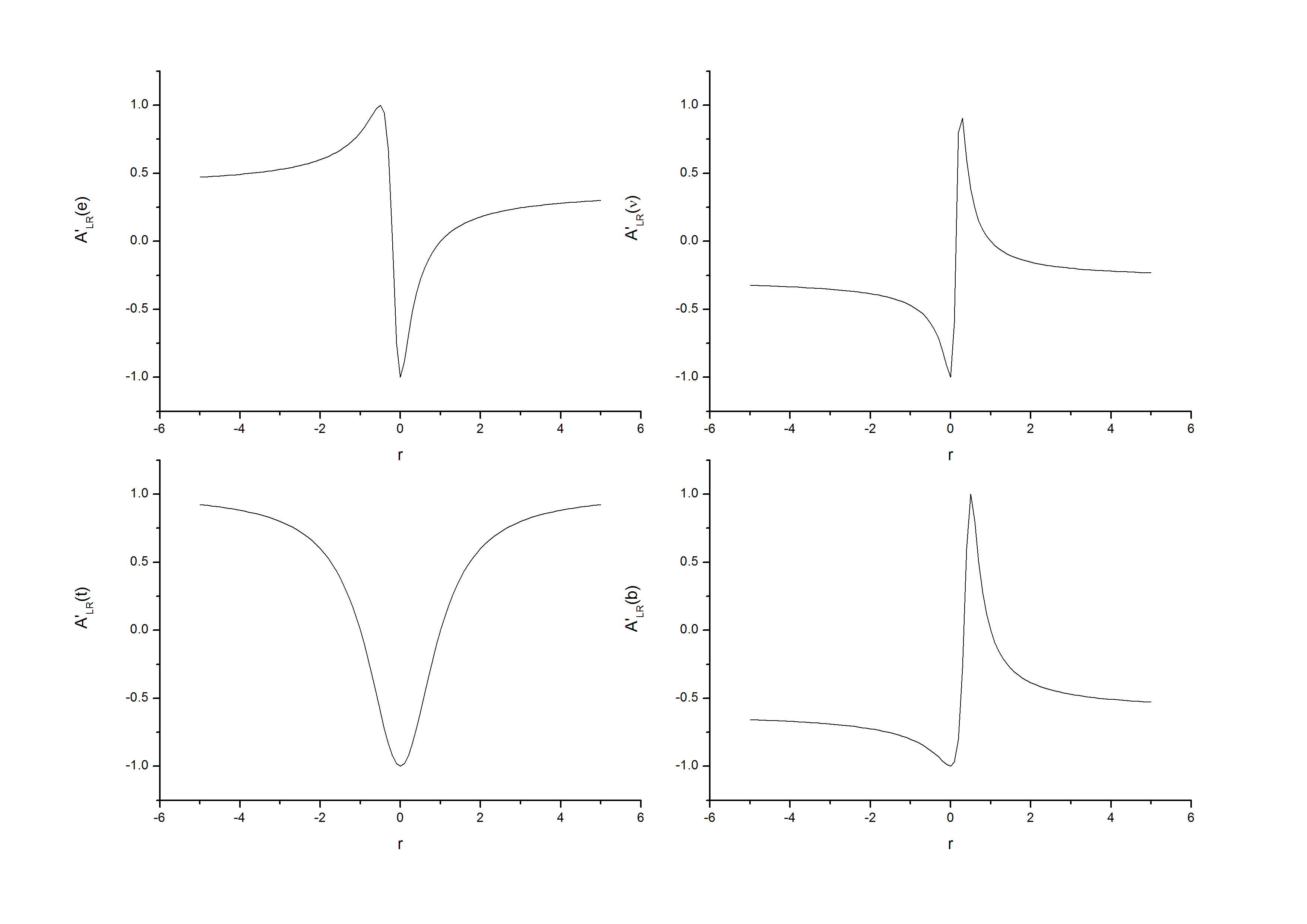}
\label{ALR}
\caption{left-right asymmetries vs. charge ratio $r$}
\end{figure}
\begin{figure}
\includegraphics[scale=0.50]{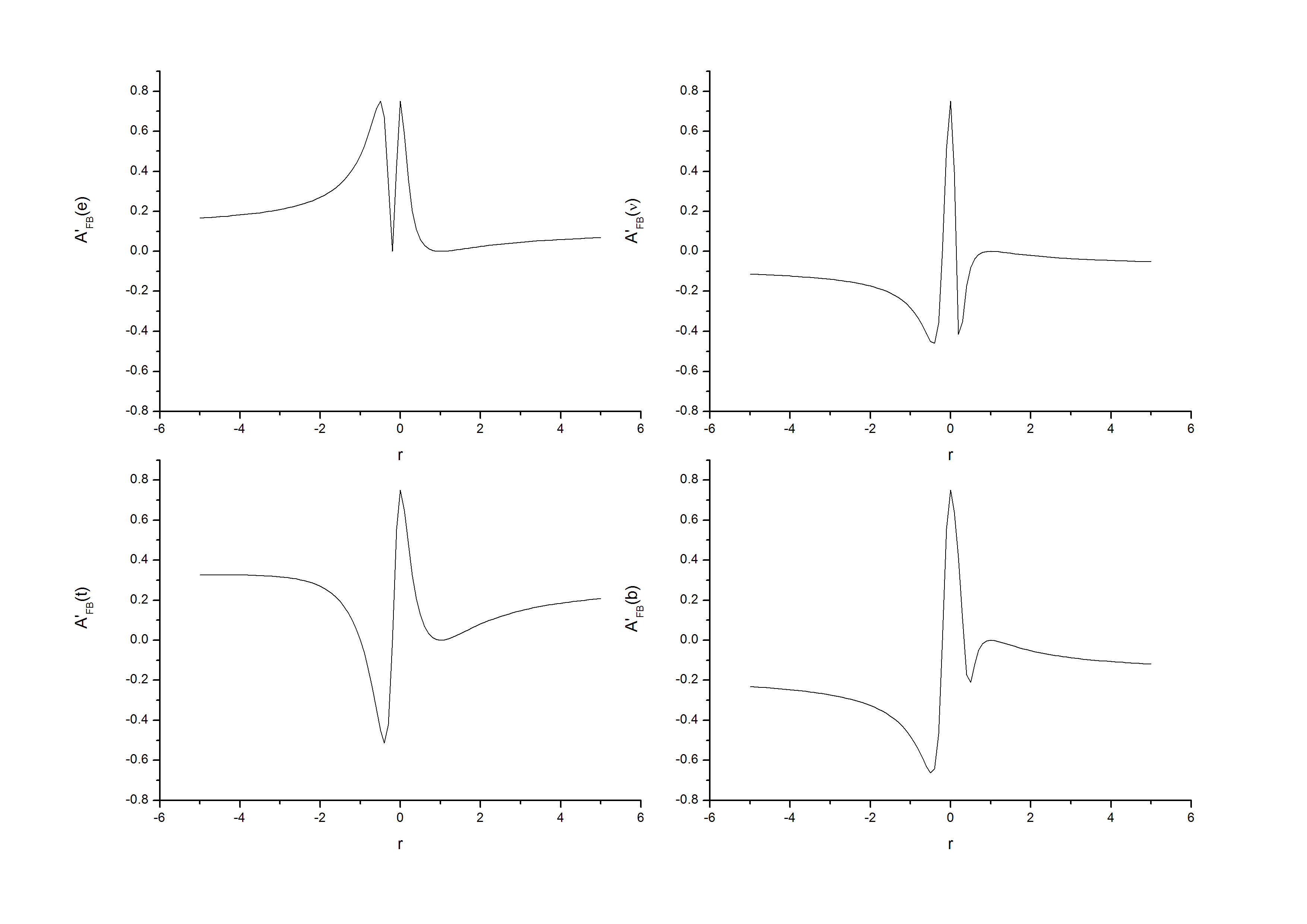}
\label{AFB}
	\caption{forward-backward asymmetries vs. charge ratio $r$}
\end{figure}

Recalling formula (\ref{DecayWidthI}), the four expressions are not completely linearly independent. These yield  a sum rule 
\begin{eqnarray}
\Gamma_{Z'}(u\bar{u})-\Gamma_{Z'}(d\bar{d})+\Gamma_{Z'}(\nu\bar{\nu})-\Gamma_{Z'}(e\bar{e})=0.
\label{sumrule1}
\end{eqnarray}
The sum rule predicts a simple relation between leading order decay widths.

In the above discussion, we have assumed charge assignments for case 1. By setting the charge ratio  $r=1/4$ in the results for case 1, we obtain the results for case 2. 
The sum rule then becomes
\begin{eqnarray}
\frac{\Gamma_{Z'}(u\bar{u})}{17}=\frac{\Gamma_{Z'}(d\bar{d})}{5}=\frac{\Gamma_{Z'}(e\bar{e})}{45}=\frac{\Gamma_{Z'}(\nu\bar{\nu})}{9}
\end{eqnarray}
Similarly, the results for case 3 correspond to setting charge ratio $r=1$ in case 1.
The sum rule becomes
\begin{eqnarray}
9\Gamma_{Z'}(u\bar{u})=9\Gamma_{Z'}(d\bar{d})=\Gamma_{Z'}(e\bar{e})=\Gamma_{Z'}(\nu\bar{\nu})
\end{eqnarray}
\section{Global Fit}\label{SECfit}
We choose for convenience the fine structure constant $\alpha$, Fermi constant $G_F$ and $Z$ boson mass $M_Z$ as our three  input fitting parameters.
We can now consider the $Z'$ corrections to these parameters. 
At the tree level, the Fermi constant keeps the same form as  in SM. 
The fine structure constant is defined by the electromagnetic coupling $\alpha=\frac{e^2}{4\pi}$. 
Using (\ref{electriccharge}), the experiment value for $\alpha$ should correspond to a normalized electromagnetic coupling with new physics effect   
\begin{eqnarray}
\alpha&=&\frac{{e^*}^2}{4\pi}.
\end{eqnarray}
In cases 1 and 2,  the renormalized electric charge has the  form
\begin{eqnarray}
e=e^*\frac{1}{1+\frac{1}{s_W}\Delta_{12}}.
\end{eqnarray}
In case 3, $e^*$ is
\begin{eqnarray}
e^*=e+\frac{e}{s_W}\Delta_{12}+g''\Delta_{32}y'_i,
\label{EC2}
\end{eqnarray}
with $B-L$ type charges $y'_{iL}=y'_{iR}=y'_i$.
The third input parameter $M_Z$ is easily isolated in the  $Z'$ correction from (\ref{ZprimeMass}). 
Thus,  an electroweak observable $\mathcal{O}_{th}(G_F,\alpha^*,M_Z^*)$ can be divided into two parts: one is $\mathcal{O}_{SM}(G_F,\alpha,M_Z)$ coming from SM fitting values in \cite{PDG2010}, and the another is $\Delta\mathcal{O}_{Z'}$ coming from  the$Z'$ new physics correction \begin{eqnarray*}
\mathcal{O}_{th}(G_F,\alpha^*,M_Z^*)=\mathcal{O}_{SM}(G_F,\alpha,M_Z)+\Delta\mathcal{O}_{Z'}.
\end{eqnarray*}

The difference between the present experimental data and SM fitting results
will provide a narrow space for the $Z'$ correction $\Delta\mathcal{O}_Z$.
We process a global fit by $\chi^2$ function
\begin{eqnarray*}
	\chi^2=\sum_i\left(\frac{\mathcal{O}^i_{exp}-\mathcal{O}^i_{th}}{\delta\mathcal{O}^i}\right)^2
=\sum_i\left(\frac{\mathcal{O}^i_{exp}-(\mathcal{O}^i_{SM}+\Delta\mathcal{O}^i_{Z'})}{\delta\mathcal{O}^i}\right)^2
\end{eqnarray*} 
with an experimental value $\mathcal{O}^i_{exp}$, an experimental error $\delta\mathcal{O}^i$, a theoretical value $\mathcal{O}^i_{th}$, SM fitting result 
$\mathcal{O}^i_{SM}$, and a new physics correction $\Delta\mathcal{O}^i_{Z'}$. Superscript $i$ indexes different observables.
Note that the weighting of an observable is larger if the standard deviation is smaller. The minimum of $\chi^2$ is defined by 
\begin{eqnarray}
	\frac{\partial}{\partial \Delta_i}\chi^2=0
\end{eqnarray}
for all $\Delta_i$. Solving the equations, we obtain the proper fitting values of the new parameters. 
The standard deviation of the new parameter is determined from the diagonal matrix elements of the error matrix.  
\section{Conclusion}\label{SECconclusion}
	
First, we fitting case 1 in which the Stueckelberg couplings $\tilde{g}'$ vanished  to maintain charge assignments more degree of freedom. The independent mixing parameters $\Delta_{11}$, $\Delta_{12}$, $\Delta_{21}$ and $y'_ug''\Delta_{31}$ can be fitted by the EWPO listed in Table \ref{tbl:fitresult}. $\Delta_{22}$ is determined by the constraint relation $s_W\Delta_{22}=c_W\Delta_{12}$. Although other mixings $\Delta_{13}$ and $\Delta_{33}$ can not be fitted directly, these can be predicted to only yield a slight shift to the electroweak observables from rotation matrix elements in Appendix \ref{APProtationmatrix}. From Table \ref{tbl:fitresult}, the typical values of the mixing parameters are of order $10^{-3}$. The shifts to EWPO are listed in Table \ref{tbl:zprimepull}. We find that the shift is not sensitive to a large charge ratio $r$ 

\begin{table}
\centering
\caption{Fit Results: center value $\Delta_{ij}$ and standard deviation $\delta_{ij}$ are in units of  $10^{-3}$. $\Delta_{31}$ can only be fitted up to a factor $g''y'_u$ and corresponding $\delta_{31}$ is global standard deviation of $g''y'_u\Delta_{31}$. We take the range of charge ratio $r$ from $-5$ to $5$ and the best value of $\chi^2$ is $20.9$}\label{tbl:fitresult}
\begin{tabular}{c|cc|cc|cc|cc}
\hline\hline
charge ratio $r$ & $\Delta_{11}$ & $\delta_{11}$ & $\Delta_{12}$ & $\delta_{12}$ & $\Delta_{21}$ & $\delta_{21}$ & $g''y'_{u}\Delta_{31}$ & $\delta_{31}$
\\
\hline
-5 & -0.35 & 0.29 & 0.12 & 0.32 &-0.81 & 0.92 & -0.010 & 0.0074
\\
-4 & -0.35 & 0.29 & 0.11 & 0.32 &-0.79 & 0.91 & -0.013 & 0.0091
\\
-3 & -0.33 & 0.28 & 0.10 & 0.32 &-0.75 & 0.90 & -0.017 & 0.012 
\\
-2 & -0.31 & 0.28 & 0.078 & 0.32 &-0.67 & 0.89 & -0.024 & 0.017 
\\
-1 & -0.24 & 0.26 & 0.017 & 0.32 &-0.47 & 0.86 & -0.043 & 0.031 
\\
-0.5 & -0.15 & 0.25 & -0.073 & 0.33 &-0.16 & 0.86 & -0.072 & 0.051
\\
0 & 0.35 & 0.42 & -0.54 & 0.52 & 1.4 & 1.5 & -0.22 & 0.15 
\\
0.1 & 0.86 & 0.67 & -1.0 & 0.73 & 3.0 & 2.2 & -0.37 & 0.24 
\\
0.25 & -0.15 & 20 & 0.051 & 19 & -0.34 & 62 & 0 & 5.8
\\
0.5 & -1.1 & 0.73 & 0.85 & 0.63 & -3.2 & 2.2 & 0.22 & 0.15 
\\
1 & 0.63 & 0.43 & 0.38 & 0.40 & -1.7 & 1.3 & 0.072 & 0.051
\\
2& -0.45 & 0.35 & 0.25 & 0.35 & -1.3 & 1.1 & 0.031 & 0.022
\\
3&-0.46 & 0.34 &0.22 & 0.34 & -1.1 & 1.0 & 0.020 & 0.014
\\
4 & -0.44 & 0.33 & 0.20 & 0.34 & -1.1 & 1.0 & 0.014 & 0.010
\\
5 & -0.43 & 0.32 & 0.19 & 0.34 & -1.0 & 1.0 & 0.011 & 0.0081
\\
\hline\hline
\end{tabular}
\end{table}

\begin{table}
\centering
\caption{$Z'$ Pull}\label{tbl:zprimepull}
   \begin{tabular}{cc|ccccccccc}
\hline\hline
   \multirow{2}*{Quantity}  & \multirow{2}*{SM Pull}  & \multicolumn{9}{c}{$Z'$ Pull at $r$}\\ \cline{3-11}
   && -2&-1& -0.5&0&0.1&0.25&0.5&1&2\\ \hline
   $M_Z$ [GeV]						& 0.1  & -0.10  & -0.10  & -0.10  &-0.10  &-0.10  &-0.10  &-0.10&-0.10&-0.10 \\ \hline
   $\Gamma_Z$ [GeV]			& -0.1  &0.26	&0.26	&0.26	&0.27	&0.27	&0.05	&0.26	&0.26	&0.26 \\ \hline
   $\Gamma_{had}$ [GeV]	& -  &0.18 &0.18 & 0.18 & 0.19 & 0.19 & 0.07&0.18 & 0.18 & 0.18\\ \hline
   $\Gamma_{inv}$ [MeV]		& -  & 0.16 & 0.16 & 0.16 & 0.16 & 0.16 &  -0.02 & 0.16 &  0.15 & 0.15 \\ \hline
   $\Gamma_{l^+l^-}$ [MeV]&  - &-0.43&	-0.43&	-0.43&	-0.44&	-0.44&	-0.01&	-0.45&	-0.44& -0.44 \\ \hline
   $\sigma_{had}$ [nb]			&  1.5 &-0.80&	-0.80&	-0.81&	-0.81&	-0.83&	-0.01&	-0.82&	-0.81&	-0.81\\ \hline
   $R_e$ 									& 1.4  &0.27&	0.27&	0.27&	0.28&	0.28&	0.04&	0.28&	0.27&	0.27\\ \hline
   $R_b$ 									& 0.8  &0.05&	0.05&	0.05&	0.05&	0.05&	-0.01&	0.05&	0.05&	0.05  \\ \hline
   $R_c$ 									& 0.0  &-0.02&	-0.02&	-0.02&	-0.02&	-0.02&	0.00&	-0.02&	-0.02&	-0.02\\ \hline
   $A^e_{FB}$ 						& -0.7  &0.06&	0.06&	0.06&	0.06&	0.06&	0.07	&0.06&	0.06&	0.06 \\ \hline
   $A^b_{FB}$ 						&  -2.7 &0.34&	0.34&	0.34&	0.34&	0.34&	0.37&	0.34&	0.34&	0.34\\ \hline
   $A^c_{FB}$ 						& -0.9  &0.13&	0.13&	0.13&	0.13&	0.13&	0.13&	0.13&	0.13&	0.13 \\ \hline
   $A^s_{FB}$ 						& -0.6  &0.05&	0.05&	0.05&	0.05&	0.05&	0.05&	0.05&	0.05&	0.05\\ \hline
   $A_e$ 									& 1.8  &0.33&	0.33&	0.33&	0.33&	0.32&	0.38&	0.33&	0.33&	0.33\\ \hline
   $A_b$ 									&  -0.6 &0.02	&0.02&	0.02&	0.02&	0.02&	0.00&	0.02&	0.02&	0.02\\ \hline
   $A_c$ 									&  0.1 &0.03&	0.03&	0.03&	0.03&	0.03&	0.01&	0.03&	0.03&	0.03 \\ \hline
   $A_s$ 									& -0.4  & 0.00&	0.00&	0.00&	0.00&	0.00&	0.00&	0.00&	0.00&	0.00 \\ \hline\hline
   \end{tabular}
\end{table}

Notwithstanding the narrow parameter space left for $Z'$, the $S$ parameter in (\ref{S}) can still take negative values from $r=-0.1$ to $r=0.28$ within $95\%$ CL (see Fig. 5 for details).
\begin{figure}
\includegraphics[scale=0.25]{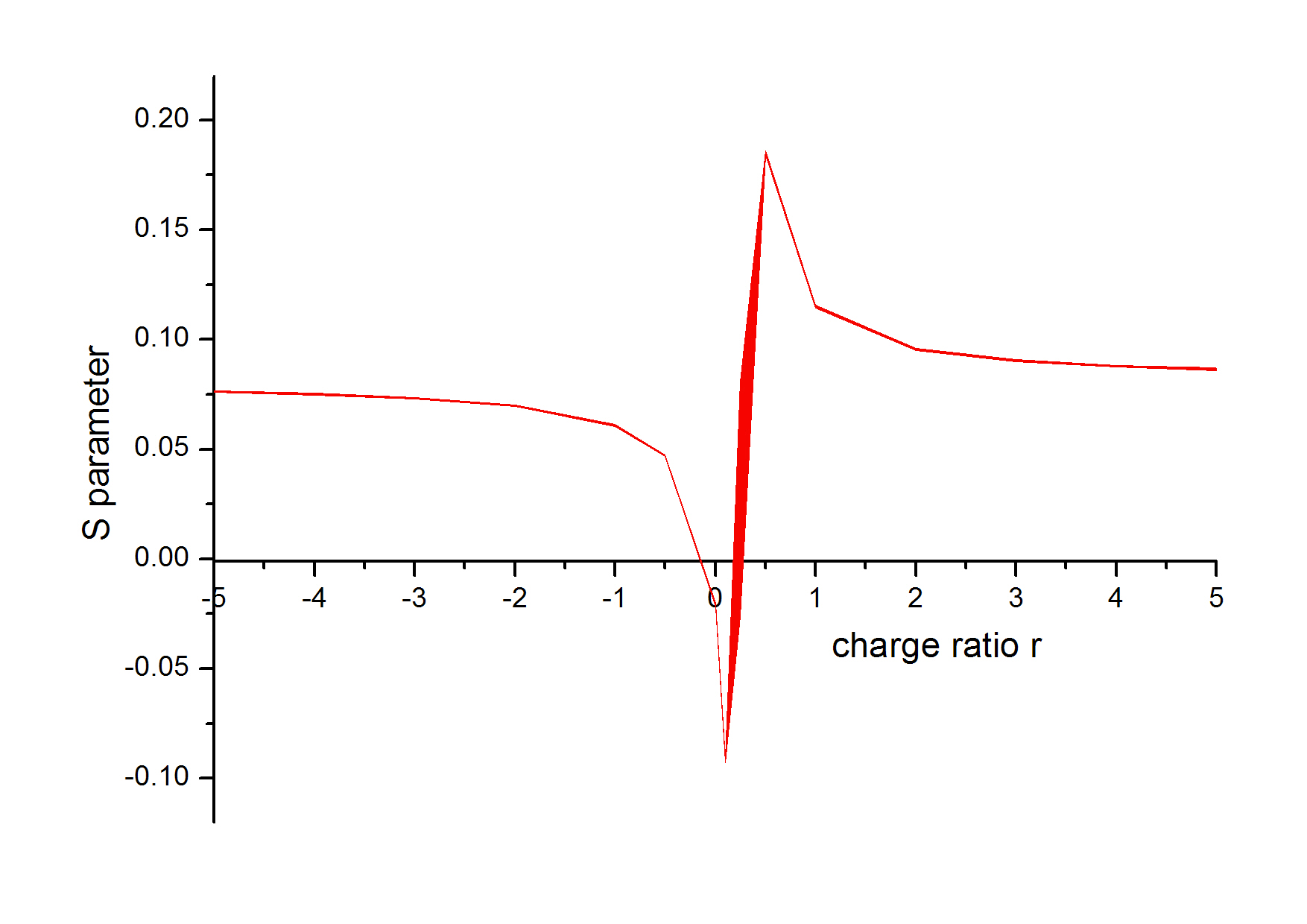}
\label{fig:S}
	\caption{$S$ parameter range vs. r within 95\% CL}
\end{figure}
A negative $S$ parameter contributing extra neutral bosons has also indicated by \cite{ErlerPRL2000,LiuZPC1994,AppelquistPRD2003}.

The $T$ parameter almost vanishes due to small $(g''\Delta_{31})^2$ in despite of a free charge $y'_u$ in (\ref{T}).

The possible range for the $U$ parameter can be calculated in terms of the fitted results in Table \ref{tbl:fitresult} (see Fig. 6 for details).  

\begin{figure}
\includegraphics[scale=0.25]{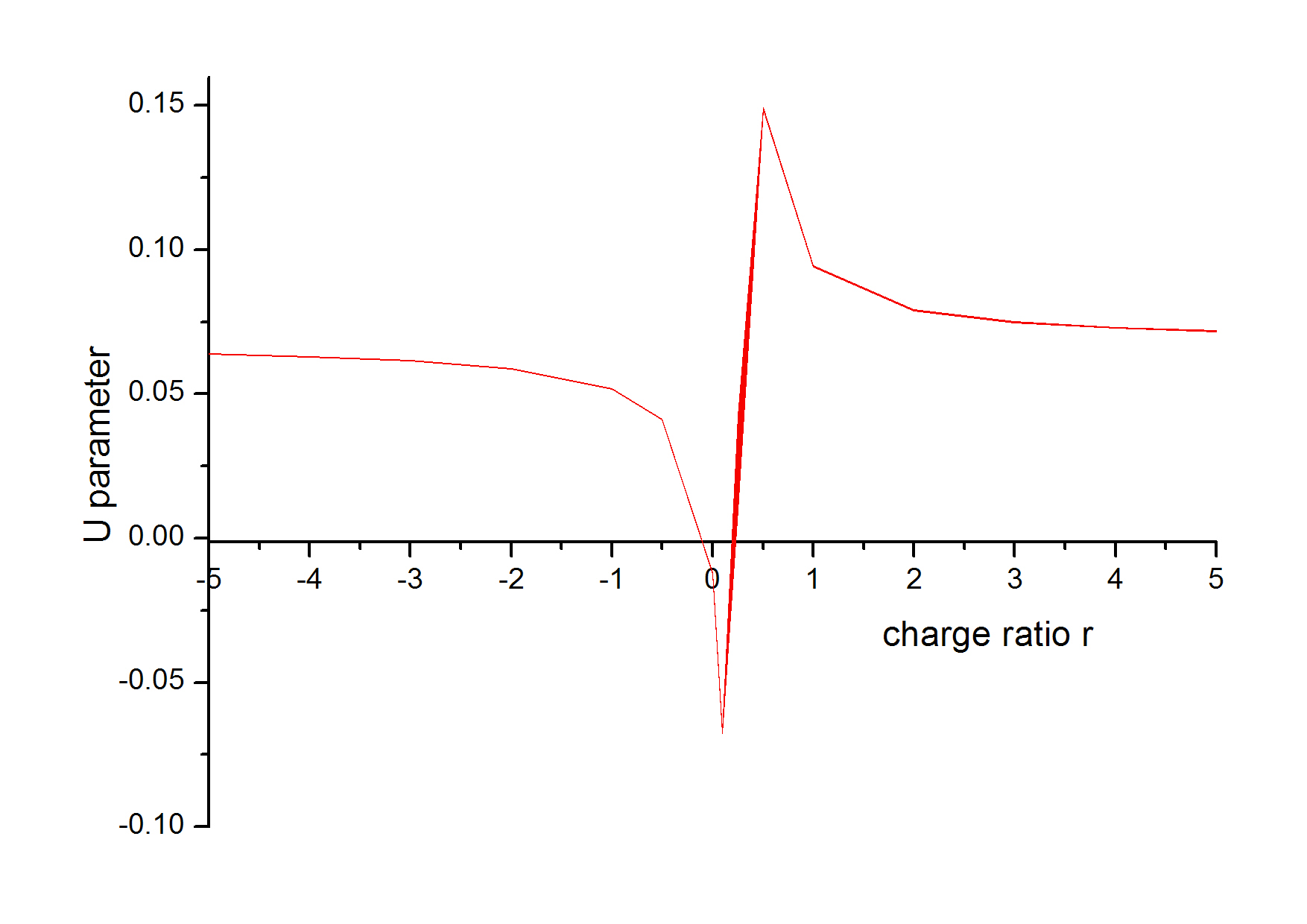}
\label{fig:U}
	\caption{$U$ parameter range vs. r within 95\% CL}
\end{figure}

When the fixed charge ratio at $r=1/4$, the above results yield those in case 2.

The third choice, case 3, correspond to a non-vanishing Stueckelberg coupling. The right-handed charge is then  equal to the left-handed charge $y'_R=y'_L$. $\Delta_{32}$ effects EWPO by shifting the normalized electric coupling in (\ref{EC2}). However, the situation is now a little different. Both mixing parameters $\Delta_{12}$ and $\Delta_{32}$ affect the electroweak physics sector in the same fashion, i.e. a shifting renormalized electric coupling. Thus, a $\chi^2$ fit only resolves the combination of $\Delta_{12}$ and $\Delta_{32}$. The fitted results are  
\begin{eqnarray}
\Delta_{11}&=&-0.00063+0.00017\Delta_{32}
\\
\Delta_{12}&=&0.00038-1.6\Delta_{32}
\\
\Delta_{21}&=&-0.0017+0.00057\Delta_{32}
\\
\Delta_{31}&=&0.000072-0.000014\Delta_{32}
\end{eqnarray}
Obviously, the above result is consistent with a vanishing Stueckelberg coupling for case 1 with $r=1$ in Table \ref{tbl:fitresult}.

All the above conclusions have treated the charge ratio $r$ as a random input parameter. We also have considered the alternative proposal of treating  $r$ as a fitting parameter; thus, $r$ can be matched to obtain an optimal value. In that case, the minimum for $\chi^2$ appears at $r=0.25$, and other mixing parameter values are the same as those in the row corresponding to $r=0.25$ in Table \ref{tbl:fitresult}. This result means that $\nu_R$ has a very faint coupling to $Z'$ ( or no $\nu_R$) and the Stueckelberg coupling $\tilde{g}'$ must vanish.

 To summarize, we have established a model-independent platform to investigate $Z'$ physics effects and parameters range restrictions by a combination of the 
 chiral effective theory with anomaly cancellation conditions without any further assumptions.
 All possible $U(1)'$ charge assignments of the fermions can be divided into three cases in terms of 
 the right-handed neutrino $\nu_R$ and the Stueckelberg coupling. 
 We fitted  the $Z'$ contribution to electroweak precise observables and obtained a narrow range of parameters. $Z'$ still contributes a negative $S$ parameter in the allowed range.
  
\section*{Acknowledgments}
This work was  supported by National  Science Foundation of China
(NSFC) under No.11005084 and No.10947152.

\begin{appendix}
\section{mixing rotation matrix in effective theory}\label{APProtationmatrix}
The rotation matrix $U$ satisfies with (\ref{Ueq}). Solving these equations, we can get the express of matrix element of $U$, which depend on coefficients in  chiral effective Lagrangian (\ref{Lagrangian}) . Up to $p^4$ order, rotation matrix elements with $\tilde{g}'$ vanishing are list in follows
\begin{eqnarray}
\Delta_{11}
&=&-s_W^3\alpha_a+(\frac{1}{2}c_W^3+c_Ws_W^2)\alpha_b
    -\frac{2gg_Z{g''}^2}{\Delta_g^2}\beta^2
\nonumber\\
\Delta_{12} &=&\frac{s_W}{c_W}\Delta_{22}
=c_Ws_W^2\alpha_a+\frac{1}{2}s_W^3\alpha_b
\nonumber\\
\Delta_{13} &=&-\frac{2gg''}{{\Delta_g}}\beta+\frac{2(g'^2-4{g''}^2)}{{\Delta_g}}\alpha_{c}+\frac{2gg'}{{\Delta_g}}\alpha_{d}
\nonumber\\
\Delta_{21}
&=&c_W^3\alpha_a+\frac{1}{2}c_W^2s_W\alpha_b
    +\frac{2g'{g''}^2g_Z}{\Delta_g^2}\beta^2
\nonumber\\
\Delta_{23}
&=&\frac{2g'g''}{{\Delta_g}}\beta+\frac{2gg'}{{\Delta_g}}\alpha_{c}+\frac{2}{{\Delta_g}}(g^2-4{g''}^2)\alpha_{d}
\nonumber\\
\Delta_{31}
&=&\frac{2g''g_Z}{{\Delta_g}}\beta+\frac{2gg_Z}{{\Delta_g}}\alpha_{c}-\frac{2g'g_Z}{{\Delta_g}}\alpha_{d}
\nonumber\\
\Delta_{32}
&=&0
\nonumber\\
\Delta_{33}
&=&-\frac{2g_Z^2{g''}^2}{\Delta_g^2}\beta^2
\nonumber
\end{eqnarray}
A general express and details computing process can be found in paper \cite{OurJHEP2008}.
\end{appendix}

\end{document}